\def\year{2021}\relax
\documentclass[letterpaper]{article} 
\usepackage{aaai21}  
\usepackage{times}  
\usepackage{helvet} 
\usepackage{courier}  
\usepackage[hyphens]{url}  
\usepackage{graphicx} 
\usepackage{natbib}  
\usepackage{float}

\usepackage{caption} 

\usepackage[group-separator={,}]{siunitx}

\usepackage{natbib}
\usepackage{amsmath,amssymb,amsfonts}
\usepackage{graphicx}
\usepackage{subcaption}
\graphicspath{ {plots/} }
\usepackage{url}
\usepackage{xcolor}
\usepackage{dirtytalk}
\usepackage{enumitem}
\usepackage{todonotes}
\newcommand{\cmmnt}[1]{}

\usepackage{pdfcomment}
\pdfcommentsetup{draft} 
\pdfcommentsetup{color={1.0 1.0 0.0}, open=true}

\title{Twitter Dataset on the Russo-Ukrainian War}

\author {
    Alexander Shevtsov,\textsuperscript{\rm 1,\rm 3}
    Christos Tzagkarakis, \textsuperscript{\rm 1}
    Despoina Antonakaki, \textsuperscript{\rm 1}
    Polyvios Pratikakis, \textsuperscript{\rm 1,\rm 3}
    Sotiris Ioannidis, \textsuperscript{\rm 2, \rm 1}\\
}
\affiliations {
    \textsuperscript{\rm 1} Institute of Computer Science, Foundation for Research and Technology - Hellas\\
    \textsuperscript{\rm 2}  School of Electrical and Computer Engineering, Technical University of Crete\\
    \textsuperscript{\rm 3} Department of Computer Science, University of Crete\\
    \{shevtsov,tzagarak,despoina,polyvios\}@ics.forth.gr,  sotiris@ece.tuc.gr
}

\begin{document}
\nocopyright

\maketitle
\begin{abstract}

On 24 February 2022, Russia invaded Ukraine, also known now as Russo-Ukrainian War. We have initiated an ongoing dataset acquisition from Twitter API. Until the day this paper was written the dataset has reached the amount of 57.3 million tweets, originating from 7.7 million users. We apply an initial volume and sentiment analysis, while the dataset can be used to further exploratory investigation towards topic analysis, hate speech, propaganda recognition, or even show potential malicious entities like botnets.  

\end{abstract}

\section{Introduction}

Twitter consists one of the most popular and widespread online 
social networks and one of the main sources of communication and dissemination in the online world. It has been used in the past to analyse crises in the political world \cite{siapera2015gazaunderattack, chibuwe2020social, chiluwa2015war}. Considering this, we have initiated a data collection from Twitter API started on 24 February 2022 when Russia invaded Ukraine, also known as the Russo-Ukrainian War. The main goal is to use this data to analyze the main trends and topics discussed in this online discourse, watch the tendencies of users, the suspension of potential malicious entities, the sentiment of the text, the hate speech, or any propaganda that may be visible through OSNs.

Up to this date, the dataset contains 57,384,192 tweets and originated from 7,744,714 users.

\subsection{Data availability}
The dataset is available on Github, at \url{https://github.com/alexdrk14/RussoUkrainianWar_Dataset}, but due to privacy restrictions applied by Twitter API, we only provide the tweet IDs. 
The research community can retrieve the Twitter dataset from Twitter API using these IDs, considering the tweets are publicly available.

\section{Methodology}
We perform some initial analysis over the collected dataset. In this section, we provide only a glimpse of our analysis, which is deployed in an online web service located here:  \url{https://alexdrk14.github.io/RussiaUkraineWar/}. Note that the plots are updated on a daily basis.

\subsection{Volume analysis}

\begin{figure}[H]
    \centering
    \includegraphics[width=70mm,scale=1.5]{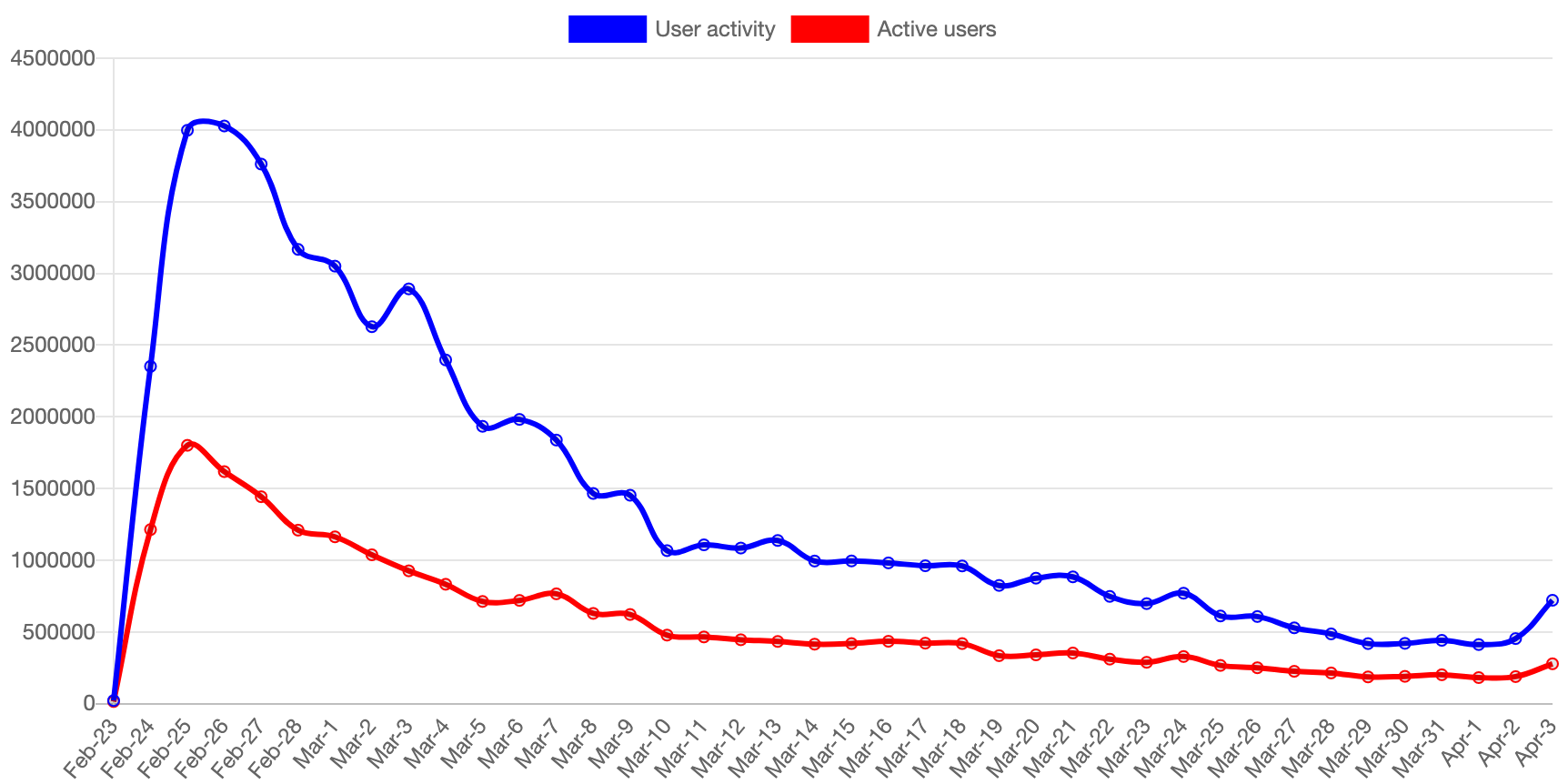}
    \caption{The Daily volume and activity of registered users.}
    \label{fig:dailyVolume}
\end{figure}

\begin{figure}[H]
    \centering
    \includegraphics[width=70mm,scale=1.5]{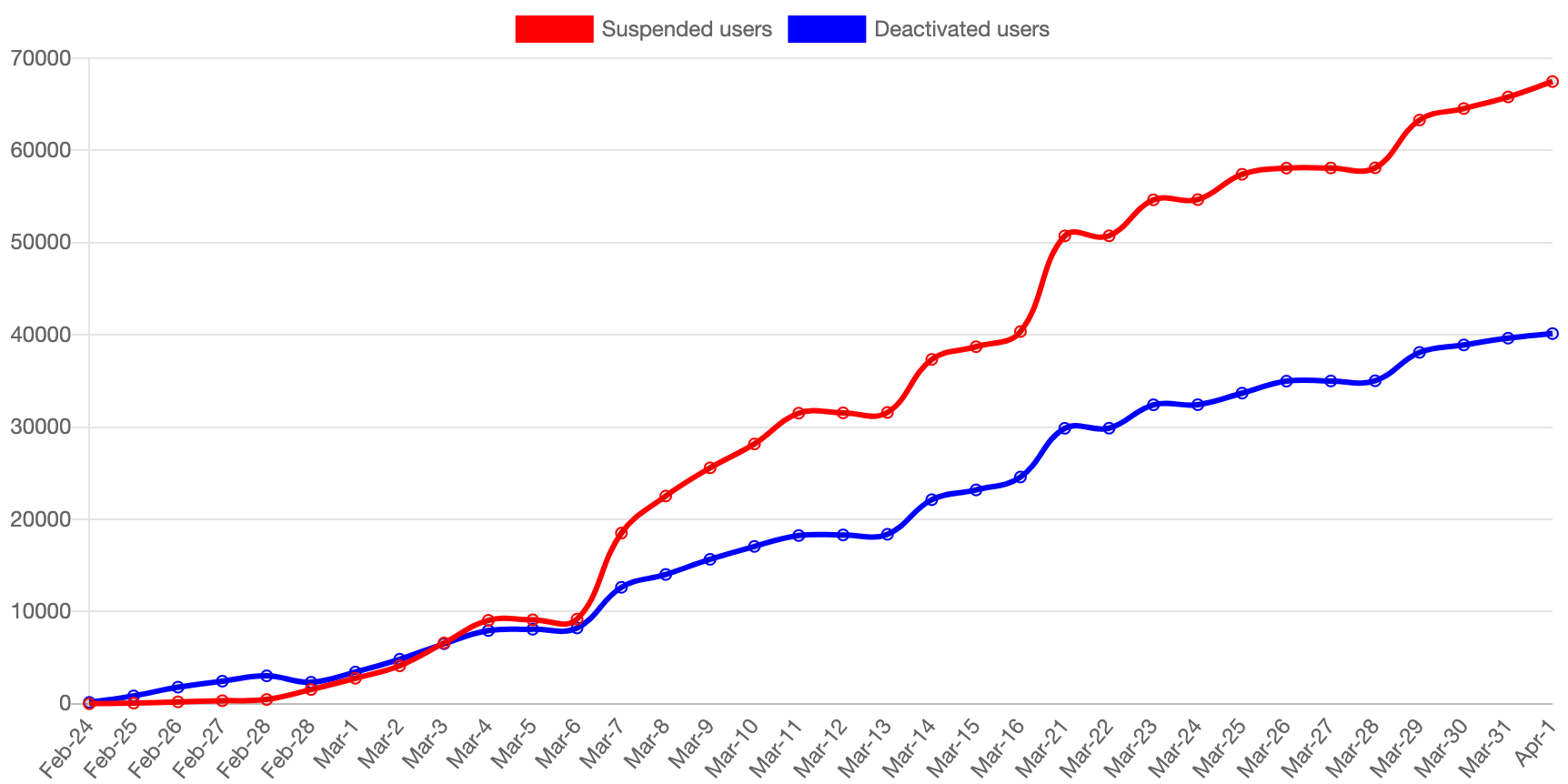}
    \caption{Daily volume of suspended/deactivated accounts.
}
    \label{fig:dailyVolumesuspended}
\end{figure}

In figure \ref{fig:dailyVolume} we present the volume of tweets per day, while in figure \ref{fig:dailyVolumesuspended} we show the volume of the suspended or deactivated accounts per day. 
We notice an increased activity in the first days of the attack, while the suspended accounts are increasing.  

\begin{figure}[H]
    \centering
    \includegraphics[width=70mm,scale=1.5]{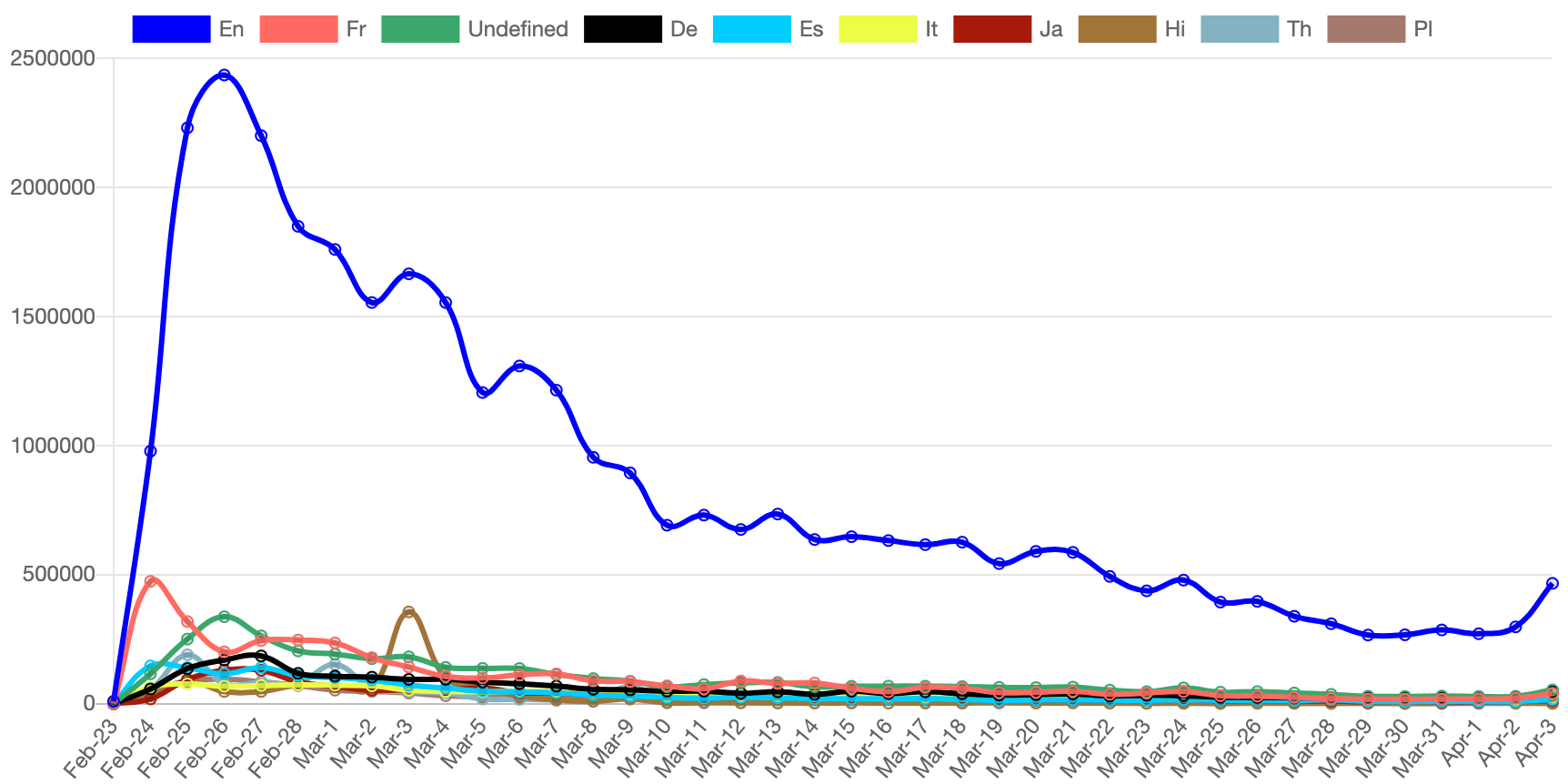}
    \caption{Daily volume of traffic based on text language.
}
    \label{fig:Dailyvolumelanguage}
\end{figure}

Figure \ref{fig:Dailyvolumelanguage} shows the daily volume of traffic based on the text language of the tweet. We only presented the ten most popular languages in the collected dataset. We notice that the language used in the majority of tweets is English.  

\begin{figure}[H]
    \centering
    \includegraphics[width=70mm,scale=1.5]{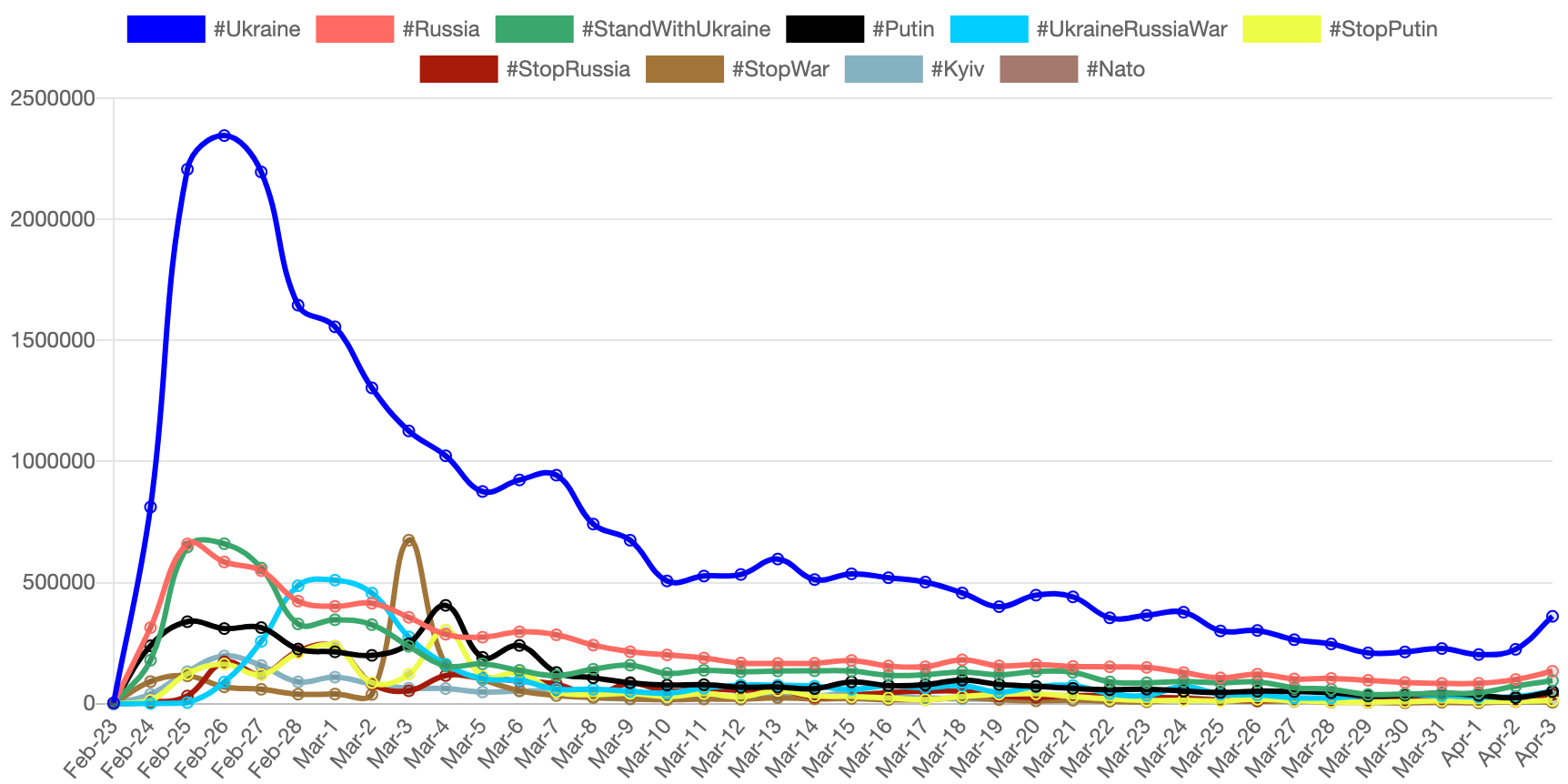}
    \caption{Daily volume of HTs traffic
}
    \label{fig:DailyvolumeHTs}
\end{figure}

Figure \ref{fig:DailyvolumeHTs} presents the daily volume of HTs traffic, while only the ten most popular hashtags in the collected dataset are included. As it seems, the most popular HT is \#Ukraine.

\begin{figure}[H]
    \centering
    \includegraphics[width=70mm,scale=1.5]{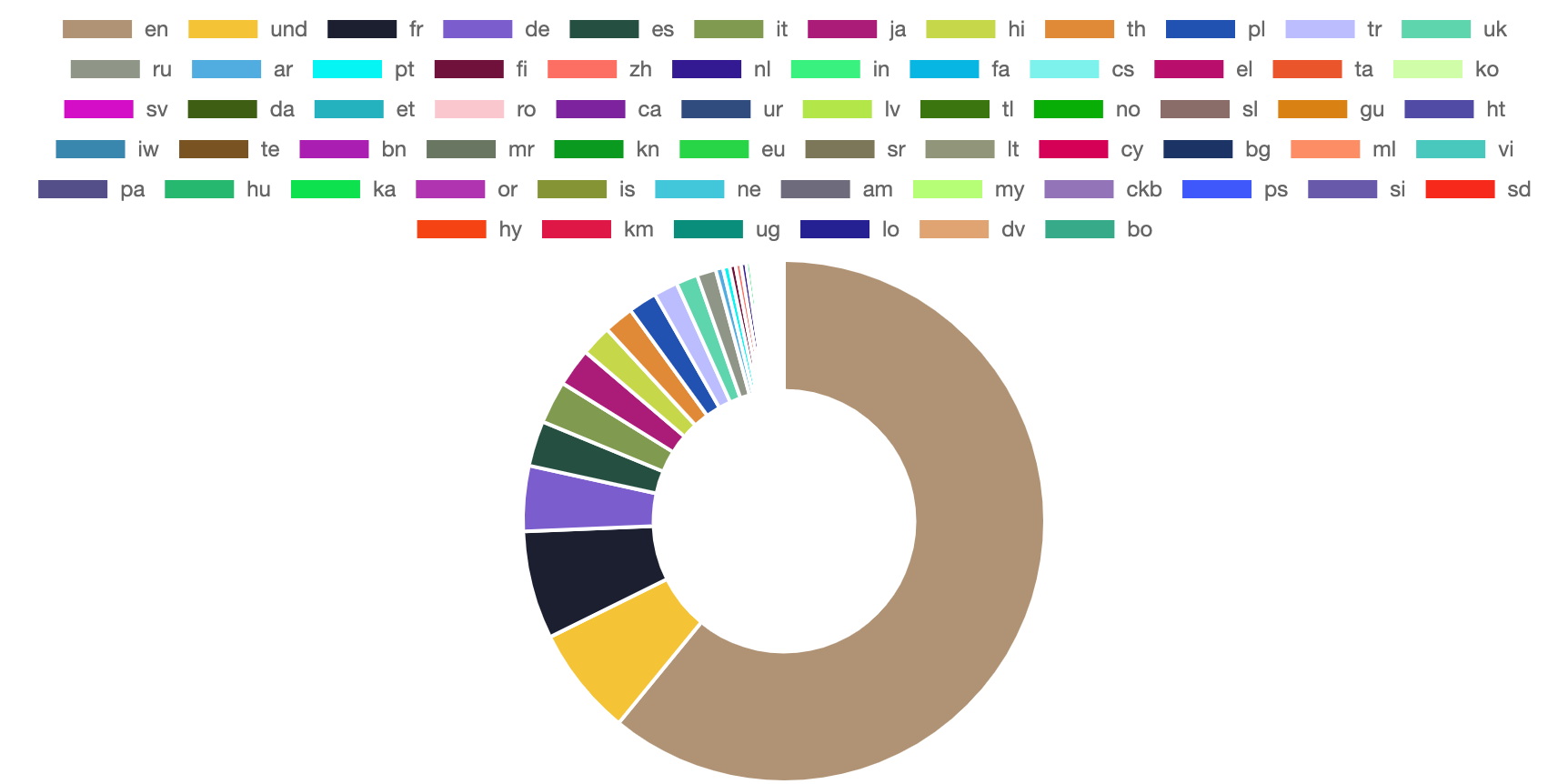}
    \caption{Total tweets' volume based on text language
}
    \label{fig:Totaltweetsvolume}
\end{figure}

Finally, in figure \ref{fig:Totaltweetsvolume} we show the total volume of tweets, based on text language, where we present all the languages in the collected dataset. As shown in figure \ref{fig:Dailyvolumelanguage}, the most popular language in tweets is English.

\subsection{Sentiment Analysis}

In this subsection, we present an initial sentiment analysis of the collected dataset by using Vader \cite{HuttoG14}. The results are shown in the figure below. 

Specifically, in figure \ref{fig:Dailypositive} we show the daily positive sentiment between Ukraine and Russia, with the higher values representing bigger support by Twitter users.

\begin{figure}[H]
    \centering
    \includegraphics[width=70mm,scale=1.5]{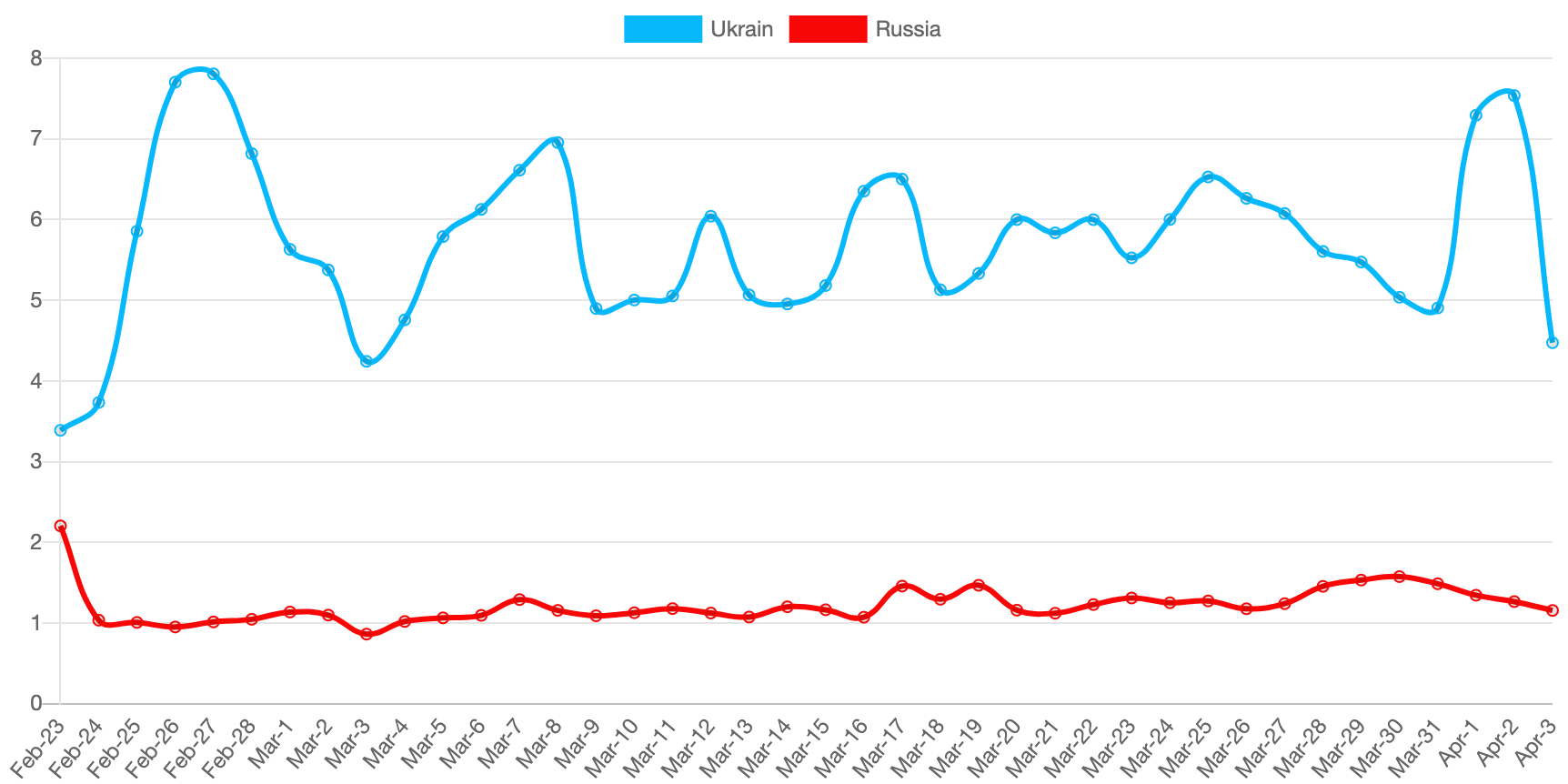}
    \caption{
Daily positive sentiment between Ukraine and Russia.
}
    \label{fig:Dailypositive}
\end{figure}

\begin{figure}[H]
    \centering
    \includegraphics[width=70mm,scale=1.5]{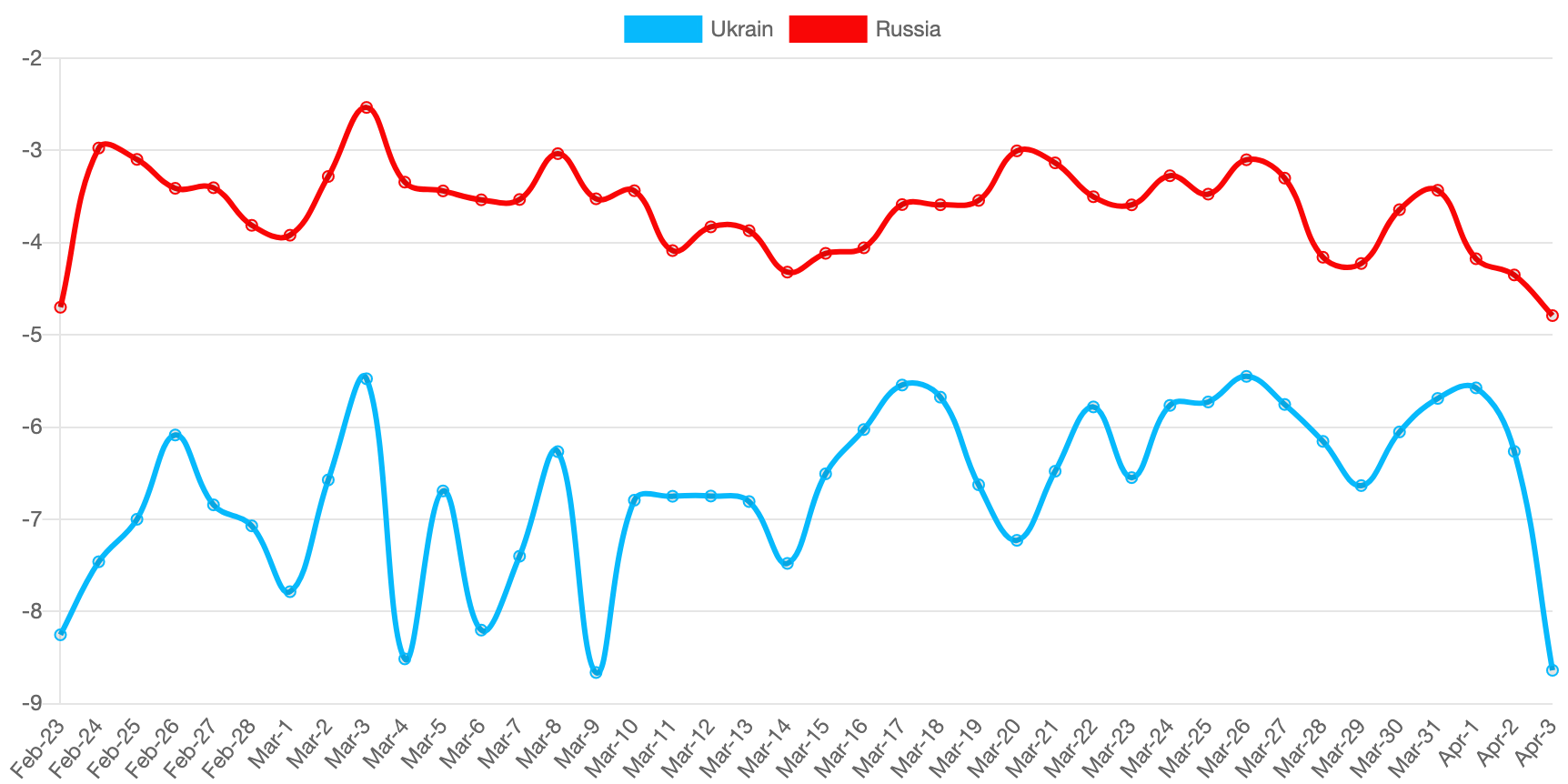}
    \caption{
Daily negative sentiment between Ukraine and Russia.
}
    \label{fig:Dailynegative}
\end{figure}

Figure \ref{fig:Dailynegative} shows the negative sentiment between Ukraine and Russia per day, with lower values corresponding to higher disagreement (sadness, rage, etc.) by Twitter users.

\begin{figure}[H]
    \centering
    \includegraphics[width=70mm,scale=1.5]{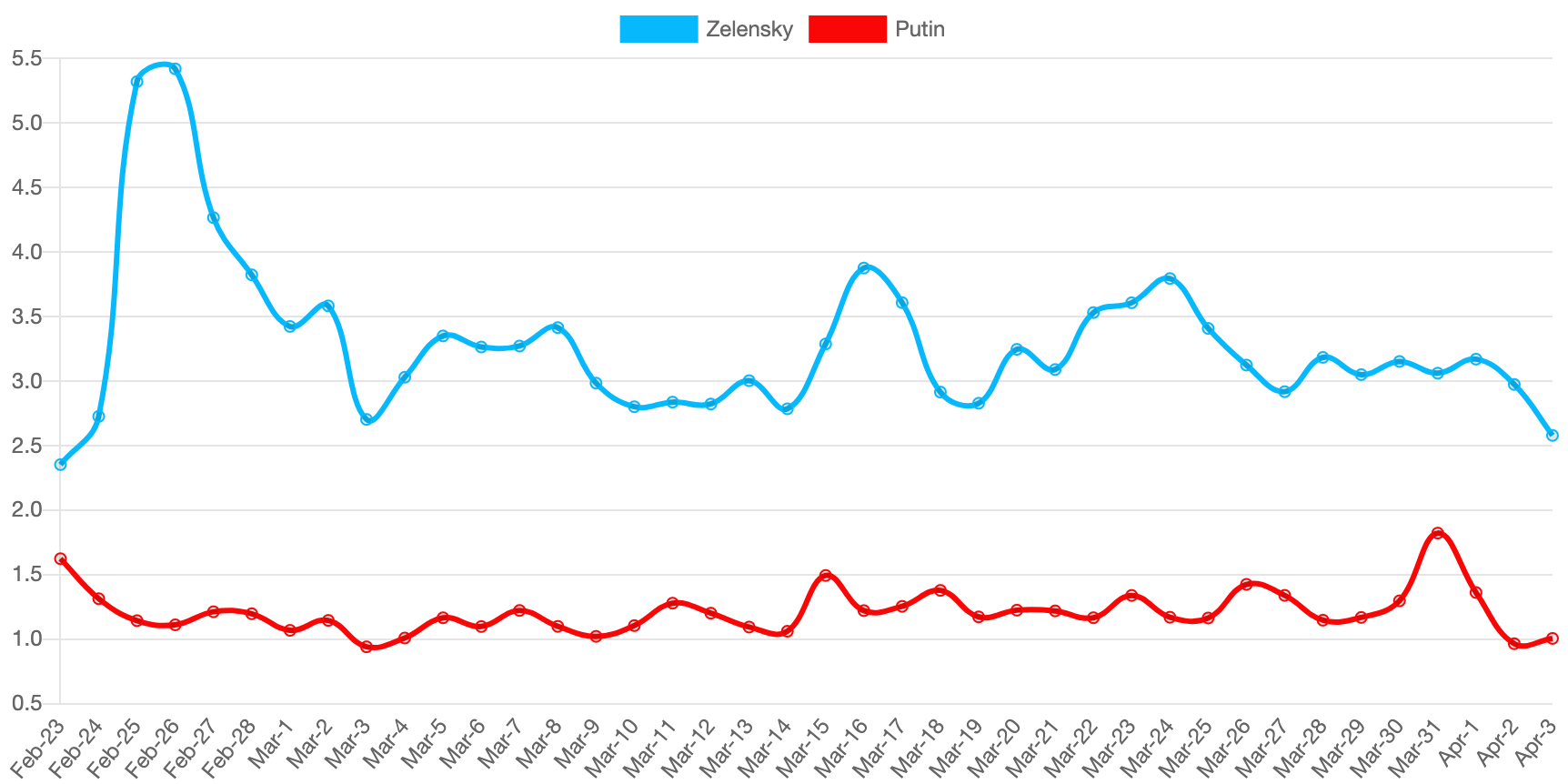}
    \caption{
Daily positive sentiment between Ukraine and Russia Presidents.
}
    \label{fig:DailypositivePresidents}
\end{figure}

\begin{figure}[H]
    \centering
    \includegraphics[width=70mm,scale=1.5]{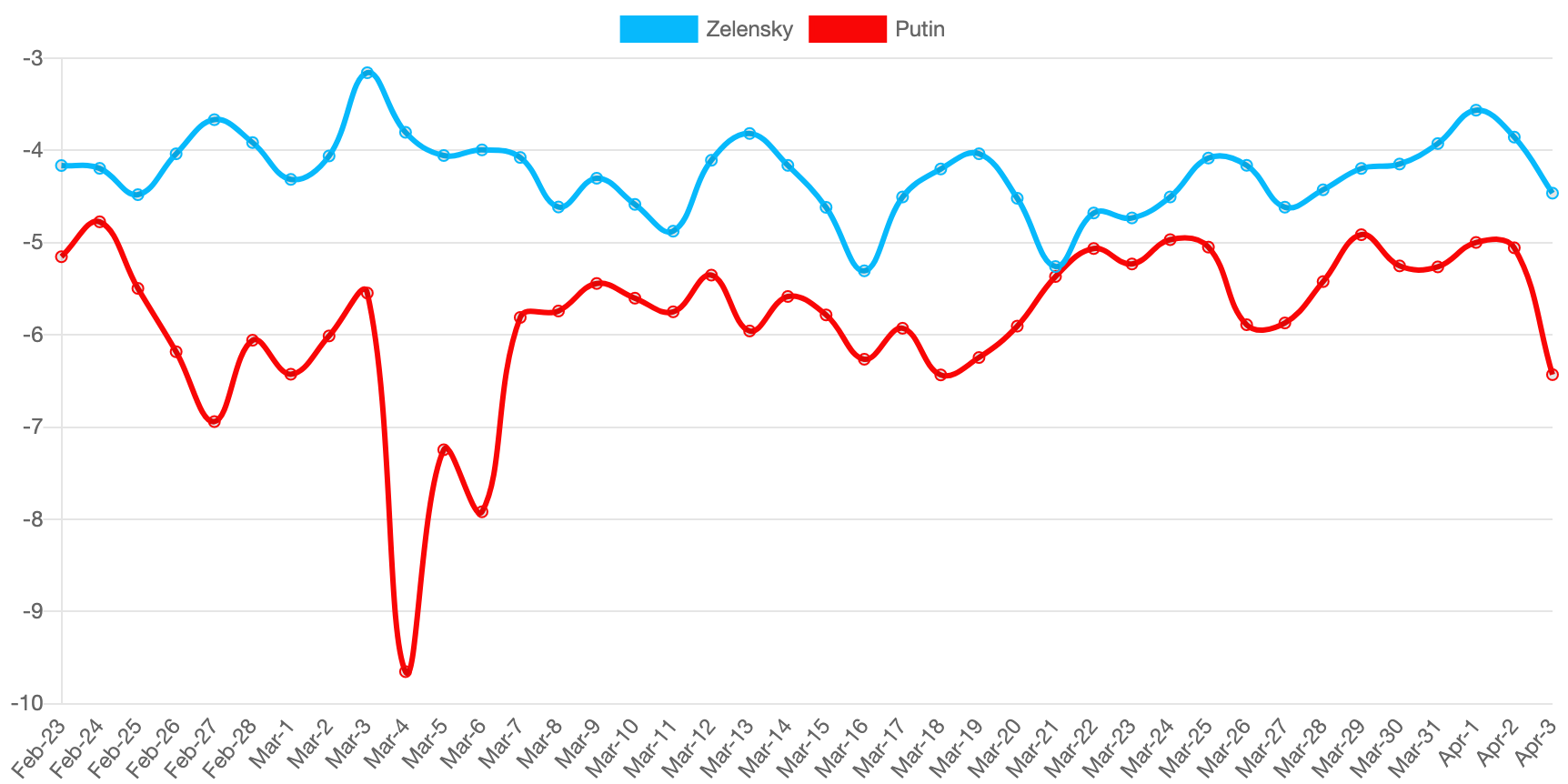}
    \caption{
Daily negative sentiment between Ukraine and Russia Presidents.
}
    \label{fig:DailynegativePresidents}
\end{figure}

Finally, we plot the positive and negative sentiment between Ukraine and Russia Presidents in figures \ref{fig:DailypositivePresidents}, \ref{fig:DailynegativePresidents} with higher values showing bigger support by Twitter users in positive sentiment while lower values show higher disagreement (sadness, rage, etc.) by Twitter users.

In table \ref{table:TopTenHS} we show the total number of tweets contained in each hashtag, for the ten most popular hashtags retrieved in our dataset, while in \ref{table:TopTenLang} we show the total number of for each language, only for the top ten most popular languages in our dataset.  

\begin{table}[]
\begin{center}
\begin{tabular}{ | l | l|  }
  \hline
  Hashtag & \# of tweets \\ \hline
 \#Ukraine & $\num[group-separator={,}]{28578739}$ \\
 \#Russia & $\num[group-separator={,}]{9070451}$\\
 \#StandWithUkraine & $\num[group-separator={,}]{6826617}$\\
 \#Putin & $\num[group-separator={,}]{4851536}$\\
 \#UkraineRussiaWar & $\num[group-separator={,}]{4007785}$\\
 \#StopRussia & $\num[group-separator={,}]{2346969}$\\
 \#StopPutin & $\num[group-separator={,}]{2332136}$\\
 \#StopWar & $\num[group-separator={,}]{1877518}$\\
 \#Kyiv & $\num[group-separator={,}]{1777401}$\\
 \#NATO & $\num[group-separator={,}]{1686092}$\\ \hline 
  \end{tabular}
\end{center}
\caption{Ten most popular hashtags in our dataset}
\label{table:TopTenHS}
\end{table}

\begin{table}[]
\begin{center}
\begin{tabular}{ | l | l|  }
  \hline
  Language & \# of tweets \\ \hline
  English & $\num[group-separator={,}]{35007332}$ \\ 
  Unclear/Mix & $\num[group-separator={,}]{3883865}$\\ 
  French & $\num[group-separator={,}]{3821685}$\\ 
  German & $\num[group-separator={,}]{2333747}$\\ 
  Spanish & $\num[group-separator={,}]{1606082}$\\ 
  Italian & $\num[group-separator={,}]{1541243}$\\ 
  Japanese & $\num[group-separator={,}]{1353848}$\\ 
  Hindi & $\num[group-separator={,}]{1076816}$\\ 
  Thai & $\num[group-separator={,}]{1047481}$\\ 
  Polish & $\num[group-separator={,}]{1012907}$\\ \hline
  \end{tabular}
\end{center}
\caption{Ten most popular text languages in our dataset}
\label{table:TopTenLang}
\end{table}

\section{Acknowledgements}
This document is the result of the research projects CONCORDIA (grant number 830927), CyberSANE (grant number 833683) and PUZZLE (grant number 883540) co-funded by the European Commission, with  (EUROPEAN COMMISSION Directorate-General Communications Networks, Content and Technology).

\bibliography{paper}

\end{document}